# Quantum mechanics without potential function

A. D. Alhaidari[a] and M. E. H. Ismail[b]

[a] *Saudi Center for Theoretical Physics, P. O. Box 32741, Jeddah 21438, Saudi Arabia*
[b] *Department of Mathematics, University of Central Florida, Orlando, FL 32816, USA*

**Abstract**: In the standard formulation of quantum mechanics, one starts by proposing a potential function that models the physical system. The potential is then inserted into the Schrödinger equation, which is solved for the wave function, bound states energy spectrum and/or scattering phase shift. In this work, however, we propose an alternative formulation in which the potential function does not appear. The aim is to obtain a set of analytically realizable systems, which is larger than in the standard formulation and may or may not be associated with any given or previously known potential functions. We start with the wavefunction, which is written as a bounded infinite sum of elements of a complete basis with polynomial coefficients that are orthogonal on an appropriate domain in the energy space. Using the asymptotic properties of these polynomials, we obtain the scattering phase shift, bound states and resonances. This formulation enables one to handle not only the well-known quantum systems but also previously untreated ones. Illustrative examples are given for two- and there-parameter systems.

## 1. INTRODUCTION

In the postulates of quantum mechanics, only two elements are specified: the state function $\Psi(t,x)$ and a single observable, the Hamiltonian operator $H$ (see, for example, [1-4]). The former determines the expectation values (measurements) of physical observables at a specific time and gives the probability density. The latter determines the time development of the system and gives its energy. Writing $H$ as the sum of the kinetic energy operator and a potential function, $H = T + V$, is a particular choice of representation that may limit the number of analytically realizable physical systems. In fact, the postulates have no reference at all to the potential function. Nonetheless, in the standard formulation of quantum mechanics, a potential function that models the physical system is the first item to be chosen. Exact solvability of the resulting Schrödinger wave equation always leads to a linear second order Sturm-Liouville type differential equation with polynomial coefficients. It is known that such an equation has polynomial solutions of all degrees if and only if the polynomial solutions are $\{x^n\}$, or Jacobi polynomials and their special cases of Hermite and Laguerre (see Chapter 20 of [5]). In the physics literature, these solutions are well known for a long time and have been tabulated and arranged into a limited number of classes by many authors (see, for example, [6] and references therein). Assigned to each class is a potential function that models the physical system such as the Coulomb, harmonic oscillator, Morse, Pöschl-Teller, Hulthén etc. In this work, however, we want to explore the possibility of realizing exactly solvable (integrable) quantum systems that may not be associated with any given or previously known potential function. This is equivalent to the search for an alternative representation of the Hamiltonian that may not be in the form $T + V$ and such that the properties of the physical system (wavefunction, scattering phase shift, energy spectrum, etc.) are obtained analytically. Instead of proposing a potential and solving the wave equation, we start by proposing a wavefunction, which is constructed as a bounded infinite sum of elements of a complete square-integrable basis set with polynomial coefficients that are orthogonal on an appropriate domain in the energy space. Using the asymptotic properties of these polynomials, we extract the scattering amplitude, phase shift, bound states and resonances. Now, the idea of using discrete square-integrable ($L^2$) basis functions to represent scattering is bold since the $L^2$ basis is usually a bound-state-like tool usually reserved for structure calculation. It is thought to be counter-intuitive for this basis to represent non-square-integrable continuum scattering states. Nonetheless, it is now an established fact that the use of complete $L^2$-basis functions proved to be very successful and efficient tool with rich mathematical underpinnings. Besides being able to handle well-known problems, our formalism enables one to present previously untreated quantum mechanical systems.

Paraphrasing, in our proposed approach to quantum mechanics, the potential function has no role to play and it does not appear in the formulation. It is only when a coordinate realization of the potential is found that one can make a correspondence with the conventional method of quantum mechanics. Such is carried out in Section 5. Moreover, our recent work in [7], where we proposed a formulation of quantum mechanics for a finite level system whose potential function is not realizable and/or analytic solution of the wave equation is not feasible, is a prelude to this one. Therefore, our formulation differs from the celebrated methods of the inverse problem in quantum mechanics and inverse scattering, where the aim is to reconstruct the potential function from the result of measurements (e.g., scattering data) [8].

Given a complete set of functions satisfying the physical boundary conditions on some configuration space, one can always propose the existence of a state function representing the physical system as a linear combination of these basis elements. This could be done without the need for prior realization of a potential function. In fact, one can use the postulates of quantum mechanics applied on this state function, which contains all information about the system [1-4]. Such wavefunction for time independent interaction could be written as $\Psi(t,x) = e^{-iEt/\hbar}\psi(E,x)$, where $E$ is the energy of the system and $x$ is the configuration space coordinate. Moreover, the time development of the system is shown explicitly and is consistent with the definition of a Hamiltonian operator as $H\Psi = i\hbar\frac{\partial}{\partial t}\Psi = E\Psi$ [1-4]. However, the Hamiltonian in this case is not required to be in the conventional format as the sum of the kinetic energy operator and a potential function since the latter is not given or analytically not realizable. Nonetheless, for a full description of the system we only need $\psi(E,x)$, which we construct as an infinite sum (linear combination) of all the discrete basis elements. That is, $\psi(E,x) = \sum_n f_n^\mu(E)\phi_n^\lambda(x)$, where $\{\phi_n^\lambda(x)\}_{n=0}^\infty$ is the complete set of $L^2$ functions that belong to the domain of the self-adjoint Hamiltonian, $\{f_n^\mu(E)\}_{n=0}^\infty$ are the expansion coefficients associated with the specific state at energy $E$, $\lambda$ is a length

–2–

scale parameter which is real and positive, and $\mu$ stands for a set of real parameters associated with the particular physical system. The conjugate state is written as $\bar{\psi}(E,x) = \sum_n f_n^\mu(E)\bar{\phi}_n^\lambda(x)$ where $\{\bar{\phi}_n^\lambda(x)\}$ is the orthogonal conjugate of $\{\phi_n^\lambda(x)\}$. That is, $\langle \phi_n^\lambda | \bar{\phi}_m^\lambda \rangle = \langle \bar{\phi}_n^\lambda | \phi_m^\lambda \rangle = \delta_{nm}$.

This paper is an expanded version of our recent letter [9] in which we introduced the formalism. Here, we provide the details, derive important formulas, present more illustrative examples and give a prospective on future lines of development based on this formulation. In Sections 2-4 below, we set up the formalism and use it to treat two- and three-parameter systems using hypergeometric polynomials that are orthogonal in the energy space. Whereas, in Section 5 we use a similar approach to handle systems in the traditional formulation of quantum mechanics using the classical orthogonal polynomials in configuration space and identify the corresponding potential functions. In paragraph (5) of Section 6, we present an analytic description of a system with resonance energies and produce figures showing its resonance structure in the complex energy plane and the wavefunction for few of the lowest order resonances.

## 2. BASES AND ENERGY POLYNOMIALS IN THE WAVEFUNCTION EXPANSION

There are several types of suitable bases for expanding the wavefunction in one-dimensional quantum mechanics.[1] To each type, we assign an appropriate and complete set of orthogonal elements using the classical orthogonal polynomials. The following are only some of the examples:

(1) The Hermite polynomial: $\phi_n^\lambda(x) \sim e^{-\lambda^2 x^2/2} H_n(\lambda x)$, where $x \in [-\infty, +\infty]$ and $H_n(z)$ is the Hermite polynomial.[2] Moreover, $\bar{\phi}_n^\lambda(x) = \phi_n^\lambda(x)$.

(2) The Jacobi polynomial: $\phi_n^\lambda(x) \sim (1-y)^{\frac{\alpha-\sigma}{2}}(1+y)^{\frac{\beta-\tau}{2}} P_n^{(\alpha,\beta)}(y)$, where $P_n^{(\alpha,\beta)}(y)$ is the Jacobi polynomial, $-1 \leq y \leq +1$ and $(\alpha,\beta) > -1$. If $x \in [-\infty, +\infty]$ then $y(\pm\infty) = \pm 1$ such as $y(x) = \tanh(\lambda x)$. Moreover, $\bar{\phi}_n^\lambda(x) = \lambda^{-1}\left(\frac{dy}{dx}\right)(1-y)^\sigma \times (1+y)^\tau \phi_n^\lambda(x)$.[3]

(3) The Laguerre polynomial: $\phi_n^\lambda(x) \sim y^{\frac{\nu-\sigma}{2}} e^{-y/2} L_n^\nu(y)$, where $L_n^\nu(y)$ is the Laguerre polynomial, $y \geq 0$ and $\nu > -1$. If, for example, $y = (\lambda x)^\tau$ with $\tau > 0$ then $x \geq 0$. On the other hand, if $y = e^{-\lambda x}$ then $x \in [-\infty, +\infty]$. Moreover, $\bar{\phi}_n^\lambda(x) = \lambda^{-1}\left(\frac{dy}{dx}\right) y^\sigma \phi_n^\lambda(x)$.

---

[1] For multi-dimensional separable system, the same applies to each dimension independently.

[2] One can also take the one-parameter basis elements: $\phi_n^\lambda(x) \sim |\lambda x|^\sigma e^{-\lambda^2 x^2/2} H_n^\sigma(\lambda x)$, where $H_n^\sigma(z)$ are the generalized Hermite polynomials and $\sigma > -\frac{1}{2}$ (see last column in Table 1).

[3] This basis is also suitable for systems on the positive real line where we can take, for example, $y = 1 - 2e^{-\lambda x}$. It is also suitable for systems confined to a finite segment of length $a$ of the real line, which could always be shifted to $x \in \left[-\frac{a}{2}, +\frac{a}{2}\right]$ where we can take, for example, $y = \sin(\pi x/a)$.

With appropriate normalization factors,[4] these elements become orthonormal (that is, $\lambda \int \bar{\phi}_n^\lambda(x)\phi_m^\lambda(x)\,dx = \delta_{nm}$) and satisfy the following completeness relation

$$\lambda \sum_n \bar{\phi}_n^\lambda(x)\phi_n^\lambda(y) = \delta(x-y). \tag{1}$$

This relation could also be obtained by evaluating the asymptotic limit of the Poisson kernel, $\lim_{n\to\infty}[K_n(x,y)]$, for the Hermite, Laguerre and Jacobi polynomials (see, for example, [10,11]). Equation (1) suggests that the wavefunction $\psi(E,x)$ is normalized in the energy as follows

$$\lambda \int_\Omega \bar{\psi}(E,x)\psi(E,y)\,d\zeta(E) = \delta(x-y), \tag{2}$$

where $d\zeta(E)$ is a suitable energy measure. Now, given a configuration space with one of the associated basis above, structural and dynamical information about any specific physical system is contained in the expansion coefficients $\{f_n^\mu(E)\}_{n=0}^\infty$, which we choose to write as

$$f_n^\mu(E) = f_0^\mu(E)P_n^\mu(\varepsilon), \tag{3}$$

where $\varepsilon$ is some proper function of $(k/\lambda)$ and $k$ is the wave number (linear momentum) associated with a particle of mass $m$ and energy $E$, which is written in the atomic units $\hbar = m = 1$ as $E = \tfrac{1}{2}k^2$. Equation (3) gives $P_0^\mu(\varepsilon) = 1$. Moreover, Eq. (2) leads to the following orthogonality relation

$$\int_\Omega \rho^\mu(\varepsilon)P_n^\mu(\varepsilon)P_m^\mu(\varepsilon)\,d\varepsilon = \delta_{nm}, \tag{4}$$

where $\rho^\mu(\varepsilon)$ is the weight function and $\Omega$ is some proper energy domain. Comparing Eq. (2) and Eq. (4), we conclude that $\rho^\mu(\varepsilon) \sim [f_0^\mu(E)]^2$. Therefore, we can take $\{P_n^\mu(\varepsilon)\}_{n=0}^\infty$ as a complete set of orthogonal polynomials. Note that if $\varepsilon = k/\lambda$ then $\Omega \in [-\infty, +\infty]$, whereas, if $\varepsilon = (k/\lambda)^2$ then $\Omega \in [0, +\infty]$. On the other hand, if $\Omega \in [-1, +1]$ then we can take, for example, $\varepsilon = \frac{(k/\lambda)^2 - 1}{(k/\lambda)^2 + 1}$ or $\varepsilon = \tanh(k/\lambda)$. Typically, there are two independent solutions to a given problem in quantum mechanics. Hence, there should exist another independent wavefunction associated with the same physical system. We write this as $\chi(E,x) = \sum_n g_n^\mu(E)\phi_n^\lambda(x)$ where $g_n^\mu(E) = g_0^\mu(E)Q_n^\mu(\varepsilon)$. Consequently, $\{Q_n^\mu(\varepsilon)\}_{n=0}^\infty$ is another set of orthogonal polynomials with $Q_0^\mu(\varepsilon) = 1$ but $Q_1^\mu(\varepsilon) \neq P_1^\mu(\varepsilon)$. The weight function associated with these new polynomials is $\omega^\mu(\varepsilon) \sim [g_0^\mu(E)]^2$. These are the polynomials of the second kind [5,12,13]. Now, if we define $\Psi_\pm(E,x) = \psi(E,x) \pm i\chi(E,x)$ then the total wavefunction will be written as $\Psi(E,x) = A_+ \Psi_+(E,x) + A_- \Psi_-(E,x)$.

---

[4] For the Hermite basis this is $\left(\sqrt{\pi}\,2^n n!\right)^{-1/2}$, and for the Laguerre basis it is $\sqrt{\Gamma(n+1)/\Gamma(n+\nu+1)}$, whereas for the Jacobi basis it reads $\sqrt{\frac{(2n+\alpha+\beta+1)\Gamma(n+\alpha+\beta+1)\Gamma(n+1)}{2^{\alpha+\beta+1}\Gamma(n+\alpha+1)\Gamma(n+\beta+1)}}$.

It is to be understood that the basis set $\{\phi_n^\lambda(x)\}$ contains only kinematical information (e.g., angular momentum, etc), which is shared by all systems whose wavefunctions are expanded as shown above. This is because the elements of this set belong to the domain of the free kinetic energy Hamiltonian and satisfy the physical boundary conditions of the corresponding configuration space. On the other hand, structural and dynamical information about the specific system under study is contained only in the energy polynomials $\{P_n^\mu(\varepsilon), Q_n^\mu(\varepsilon)\}$ and their corresponding weight functions.

### 3. ASYMPTOTICS AND SCATTERING

For bound states, the physical boundary conditions dictate that: $\lim_{|x|\to\infty} \psi(E,x) = 0$. On the other hand, for unbounded systems the wavefunction in this limit represents a free particle, which is a combination of outgoing and incoming plane waves (sinusoidal waves with some phase angle difference). That is, in the limit as $x \to \infty$,

$$\Psi(E,x) \approx A_+ e^{ikx} + A_- e^{-ikx} = |A_+| e^{+i(kx+\delta)} + |A_-| e^{-i(kx+\delta)}$$
$$= |A_-| e^{-i\delta} \left[ e^{-ikx} + R e^{2i\delta} e^{+ikx} \right] \quad (5)$$

where $R = |A_+/A_-|$ is the reflection coefficient (for elastic scattering $R = 1$) and $\delta(E)$ is the scattering phase shift, which is a measure of the phase difference between the outgoing and incoming plane waves due to the interaction in the interior region of configuration space. In the absence of interaction, it is an integer multiple of $\pi/2$ independent of the energy. The term "scattering matrix" is usually given to the term $Re^{2i\delta}$. In earlier findings [14,15], it was establish that

$$\lim_{x\to\infty} \Psi_\pm(E,x) = \lim_{N\to\infty} \sum_{n=N}^{\infty} \left[ f_n^\mu(E) \pm i g_n^\mu(E) \right] \phi_n^\lambda(x). \quad (6)$$

That is, the nature of the physical system in the asymptotic region of configuration space (with coordinate $x$) is equivalent to that which is represented by the asymptotics in function space (indexed by $n$).

In three-dimensional configuration space with spherical symmetry, radial coordinate $r$, and angular momentum quantum number $\ell$,

$$\lim_{r\to\infty} \psi(E,r) = \mathcal{A} \cos\left(kr - \ell\tfrac{\pi}{2} + \delta\right), \qquad \ell = 0,1,2,... \quad (7a)$$

In two-dimensions with cylindrical symmetry, radial coordinate $r$, and azimuthal quantum number $m$,

$$\lim_{r\to\infty} \psi(E,r) = \mathcal{A} \cos\left(kr - |m|\tfrac{\pi}{2} + \delta\right), \qquad m = 0, \pm 1, \pm 2, ... \quad (7b)$$

In one dimension with left-right symmetry[5]:

$$\lim_{|x|\to\infty} \psi(E,x) = \mathcal{A} \cos(kx + \delta), \quad (7c)$$

---

[5] The phase shift angle $\delta$ is related to the transmission ($T$) and reflection ($R$) amplitudes. These are defined in the asymptotics of the wavefunction (for normalized flux incident from left) as $\lim_{x\to+\infty} \Psi(E,x) = T e^{ikx}$, $\lim_{x\to-\infty} \Psi(E,x) = e^{ikx} + R e^{-ikx}$ and could be written for left-right symmetry as $T = \tfrac{1}{2}\left(e^{2i\theta_+} + e^{2i\theta_-}\right)$, $R = \tfrac{1}{2}\left(e^{2i\theta_+} - e^{2i\theta_-}\right)$ where $\theta_\pm$ are real angular parameters and $|R|^2 + |T|^2 = 1$.

Therefore, for a given choice of basis, $\{\phi_n^\lambda(x)\}$, physical requirements dictate that we consider only the class of orthogonal polynomials $\{P_n^\mu(\varepsilon), Q_n^\mu(\varepsilon)\}$ that satisfy the following asymptotic limit

$$\lim_{N \to \infty} \sum_{n=N}^{\infty} \left[ \sqrt{\rho^\mu(\varepsilon)} P_n^\mu(\varepsilon) \pm i\sqrt{\omega^\mu(\varepsilon)} Q_n^\mu(\varepsilon) \right] \phi_n^\lambda(x) \qquad (8)$$
$$= A\left[ \cos(kx + \Delta) \pm i\sin(kx + \Delta) \right]$$

The phase angle $\Delta$ contains the scattering phase shift $\delta$ as well as the kinematic constants like $\ell$ and $m$, which appear as part of the parameters of the basis in $\phi_n^\lambda(x)$. Requirement (8) translates into the following asymptotic properties for the orthogonal polynomials of the first and second kind in the energy[6]

$$P_n^\mu(\varepsilon) \approx A(\varepsilon) \cos\left[ n^\xi \theta(\varepsilon) + \delta(\varepsilon) \right], \qquad \text{as } n \to \infty. \qquad (9a)$$

$$Q_n^\mu(\varepsilon) \approx \pm A(\varepsilon) \sin\left[ n^\xi \theta(\varepsilon) + \delta(\varepsilon) \right], \qquad \text{as } n \to \infty. \qquad (9b)$$

where $A(\varepsilon)$ is the scattering amplitude and $\xi$ is a real positive constant that depends on the particular energy polynomial. The scattering phase shift $\delta(\varepsilon)$ depends on the energy as well as the set of interaction parameters $\{\mu\}$ attached to the polynomials $\{P_n^\mu(\varepsilon), Q_n^\mu(\varepsilon)\}$. On the other hand, bound states (if they exist) occur at discrete *real* energies $\{\varepsilon_n\}$ at which $A(\varepsilon_n) = 0$ forcing the asymptotic wavefunction to vanish thus confining the system in configuration space. However, if $A(\varepsilon_n) = 0$ at *complex* energies $\{\varepsilon_n\}$ with negative imaginary parts, then these are the resonance energies and the imaginary part forces the wavefunction to vanish with time due to the factor $e^{-iE_n t/\hbar}$. The number of bound states and resonances could be finite or infinite. Finally, for a full description of the physical system, we end up with three possible configurations:

(1) *Only continuum scattering states*: The wavefunction at energy $\varepsilon$ is written as $\psi(\varepsilon, x) = \sqrt{\rho^\mu(\varepsilon)} \sum_{n=0}^{\infty} P_n^\mu(\varepsilon) \phi_n^\lambda(x)$.

(2) *Only discrete bound states*: The $m^{\text{th}}$ bound state wavefunction (with energy $\varepsilon_m$) is written as $\psi_m(x) = \sqrt{\rho_m^\mu} \sum_{n=0}^{N} R_n^\mu(\varepsilon_m) \phi_n^\lambda(x)$, where $\{R_n^\mu(\varepsilon_m)\}$ are polynomials with the discrete orthogonality relation $\sum_{m=0}^{N} \rho_m^\mu R_n^\mu(\varepsilon_m) R_{n'}^\mu(\varepsilon_m) = \delta_{n,n'}$, and $N$ is either finite or infinite.

(3) *Both continuum as well as discrete states*: The total wavefunction is $\psi_m(\varepsilon, x) = \sqrt{\rho^\mu(\varepsilon)} \sum_{n=0}^{\infty} P_n^\mu(\varepsilon) \phi_n^\lambda(x) + \sqrt{\rho_m^\mu} \sum_{n=0}^{N} R_n^\mu(\varepsilon_m) \phi_n^\lambda(x)$, where the orthogonality relation becomes $\int_\Omega \rho^\mu(\varepsilon) P_n^\mu(\varepsilon) P_{n'}^\mu(\varepsilon) d\varepsilon + \sum_m \rho_m^\mu R_n^\mu(\varepsilon_m) R_{n'}^\mu(\varepsilon_m) = \delta_{n,n'}$.

Geronimo [16] and Case and Geronimo [17,18] explained the connection between scattering theory and the asymptotics of orthogonal polynomials. Now, the asymptotic behavior (9) is typical for Legendre, Ultra-spherical or Jacobi polynomials (see Chapter 8 in Szegő [13]). There were two attempts to extend this to general orthogonal polynomials. The first was the Szegő theory, which proved (9) when the polynomials are orthogonal with respect to a weight function $\rho(\varepsilon)$ on $[-1,+1]$ and $\int_{-1}^{+1} \ln[\rho(\varepsilon)] \times (1-\varepsilon^2)^{-1/2} d\varepsilon$ is finite [13]. The second was developed by Freud and completed by Nevai, where the starting point is the three-term recurrence relation

$$\varepsilon P_n(\varepsilon) = a_n P_n(\varepsilon) + b_{n-1} P_{n-1}(\varepsilon) + b_n P_{n+1}(\varepsilon), \tag{10}$$

and $(b_n, a_n) \to (1/2, 0)$ as $n \to \infty$ (see [19] and [20]). In the following Section, we give illustrative examples to demonstrate how one can apply this formalism to obtain physical information about the system.

---

[6] Precisely, $\delta(\varepsilon)$ in Eqs. (9) should be replaced by $\delta(\varepsilon) + C(n)$ where $C(n) = o(n^\xi)$ as $n \to \infty$.

## 4. DESCRIPTIVE EXAMPLES

As illustration, we consider quantum systems on the positive real line $s \geq 0$, where the configuration space coordinate $x$ is related to the variable $s$ by an appropriate function $s(\lambda x)$. In addition to one-dimensional systems, this also includes the radial component of spherically symmetric three-dimensional systems like the Coulomb and the spherical oscillator problems. Thus, we construct wavefunctions of the form

$$\psi(E, x) = \sum_{n=0}^{\infty} \sqrt{\rho^\mu(\varepsilon)} P_n^\mu(\varepsilon) \phi_n^\lambda(x), \tag{11}$$

where $\phi_n^\lambda(x) = \sqrt{\frac{\Gamma(n+1)}{\Gamma(n+\nu+1)}} s^{\frac{\nu-\sigma}{2}} e^{-s/2} L_n^\nu(s)$ and $\bar{\phi}_n^\lambda(x) = \lambda^{-1}\left(\frac{ds}{dx}\right) s^\sigma \phi_n^\lambda(x)$. Now, we choose polynomials $P_n^\mu(\varepsilon)$ that are orthogonal on a given energy domain $\Omega$ and find the corresponding phase shift, scattering amplitude and energy spectrum as explained above. Such a system may or may not correspond to any of the familiar quantum systems associated with known analytic potential functions. Here, we study two classes of exactly solvable problems. The first is a two-parameter class associated with the Meixner-Pollaczek polynomial and the second is a three-parameter class associated with the continuous dual Hahn orthogonal polynomial. A four-parameter class associated with the Wilson polynomial will be left for a future study.

### 4.1 The Meixner-Pollaczek polynomial class:
The energy polynomials in this case are the orthonormal version of the Meixner-Pollaczek polynomials [21], which are written as

$$P_n^\mu(y;\theta) = \sqrt{\frac{\Gamma(n+2\mu)}{\Gamma(2\mu)\Gamma(n+1)}} e^{in\theta} {}_2F_1\left(\begin{matrix}-n, \mu+iy\\ 2\mu\end{matrix}\bigg|1-e^{-2i\theta}\right), \tag{12}$$

where $y \in [-\infty, +\infty]$, $\mu > 0$ and $0 < \theta < \pi$. These polynomials satisfy the following symmetric three-term recursion relation

$$(y \sin\theta) P_n^\mu = -\left[(n+\mu)\cos\theta\right] P_n^\mu + \frac{1}{2}\sqrt{n(n+2\mu-1)} P_{n-1}^\mu \\ + \frac{1}{2}\sqrt{(n+1)(n+2\mu)} P_{n+1}^\mu \tag{13}$$

The associated normalized weight function is

$$\rho^\mu(y;\theta) = \frac{1}{2\pi\Gamma(2\mu)}(2\sin\theta)^{2\mu} e^{(2\theta-\pi)y} |\Gamma(\mu+iy)|^2. \tag{14}$$

The generating function for these polynomials is written as

$$\sum_{n=0}^{\infty} \tilde{P}_n^\mu(y,\theta) t^n = \left(1-te^{i\theta}\right)^{-\mu+iy}\left(1-te^{-i\theta}\right)^{-\mu-iy}, \tag{15}$$

–7–

where $\tilde{P}_n^\mu(y;\theta) = \sqrt{\frac{\Gamma(n+2\mu)}{\Gamma(2\mu)\Gamma(n+1)}} P_n^\mu(y;\theta)$. To derive the large $n$ asymptotic formula we use Darboux's method (see Chapter 8 of [22]). The details of the calculation are given in the Appendix where the relevant result is Eq. (A4). Since $\ln n \approx o(n)$ as $n \to \infty$ then the $n$-dependent term $y \ln(2n \sin\theta)$ in the argument of the cosine in (A4) could be ignored relative to $n\theta$. Comparing the result with Eq. (9a) gives the following scattering phase shift (modulo an integer multiple of $\pi/2$)

$$\delta(\varepsilon) = \arg\Gamma(\mu + iy) + \mu(\theta - \pi/2). \tag{16}$$

If we choose the parameters $\mu = \ell + 1$, $\theta = \pi/2$, $y = Z/k$, we obtain the following phase shift

$$\delta(E) = \arg\Gamma(\ell + 1 + iZ/k), \tag{17}$$

which is the Coulomb scattering phase shift associated with an electric charge $Z$. On the other hand, if we choose $\mu = \ell + \lambda^{-1}\sqrt{V_0}$, $\theta = \pi/2$ and $y = \sqrt{\beta^{-1}\ln(1+\lambda^2/k^2)}$, where $\beta$ and $V_0$ are real positive parameter then we obtain the associated scattering phase shift

$$\delta(E) = \arg\Gamma\left[\ell + \lambda^{-1}\sqrt{V_0} + i\sqrt{\beta^{-1}\ln(1+\lambda^2/k^2)}\right]. \tag{18}$$

This does not correspond to any of the well-known quantum systems and, to the best of our knowledge, was not treated before in the physics literature.

Now, bound states correspond to the condition (A6), which gives the energy spectrum formula $y^2 = -(n+\mu)^2$. An alternative approach to reach the same conclusion is as follows. Since bound states correspond to asymptotically zero norm states, then a simple way to generate such states is by the replacement $\theta \to i\theta$ because then the oscillatory factor $e^{in\theta}$ in (12) changes into a decaying exponential that tends to zero as $n \to \infty$. Making this replacement in the three-term recursion relation (13) and writing $\cosh\theta = a/b$ (with $a \geq b$) then multiplying both sides by $b$ and taking the limit $b \to 0$, changes the tridiagonal representation into a diagonal one that reads: $iy = -(n+\mu)$, resulting in the same energy spectrum formula. Thus, the $m^{th}$ bound state wavefunction is obtained by substituting $i(m+\mu)$ for $y$ and $i\theta$ for $\theta$ in the polynomial definition (12) giving $P_n^\mu(i(m+\mu);i\theta) = \sqrt{\frac{\Gamma(n+2\mu)}{\Gamma(2\mu)\Gamma(n+1)}} e^{-n\theta} {}_2F_1\left(\begin{matrix}-n,-m\\2\mu\end{matrix}\middle|1-e^{2\theta}\right)$. This is the discrete Meixner polynomial, which we write as $M_n^\alpha(m;\beta) = \sqrt{\frac{\Gamma(n+\alpha)\beta^n}{\Gamma(\alpha)\Gamma(n+1)}} {}_2F_1\left(\begin{matrix}-n,-m\\\alpha\end{matrix}\middle|1-\beta^{-1}\right)$, where $0 < \beta < 1$. These polynomials satisfy a discrete orthogonality relation that reads as follows (Eq. 1.9.1 and Eq. 1.9.2 in [21])

$$\sum_{m=0}^\infty \rho_m^\alpha(\beta) M_n^\alpha(m;\beta) M_{n'}^\alpha(m;\beta) = \delta_{n,n'}, \tag{19}$$

where the normalized discrete weight is $\rho_m^\alpha(\beta) = (1-\beta)^\alpha \frac{\Gamma(m+\alpha)\beta^m}{\Gamma(\alpha)\Gamma(m+1)}$. Thus, the $m^{th}$ bound state wavefunction becomes

$$\psi_m(x) = (1-e^{-2\theta})^\mu e^{-m\theta} \sqrt{\frac{\Gamma(m+2\mu)}{\Gamma(2\mu)\Gamma(m+1)}} \sum_{n=0}^\infty M_n^{2\mu}(m;e^{-2\theta}) \phi_n^\lambda(x). \tag{20}$$

Using the fact that $_2F_1\left({a,b \atop c}\big|z\right)$ is invariant under the exchange $a \leftrightarrow b$ and $\langle\phi_n^\lambda|\bar{\phi}_{n'}^\lambda\rangle = \delta_{n,n'}$, one can show that these bound states are orthonormal; that is, $\langle\psi_m|\bar{\psi}_{m'}\rangle = \delta_{m,m'}$. Now, substituting the Coulomb problem parameters given above into the energy spectrum formula $y^2 = -(n+\mu)^2$, we obtain the well-known result

$$E_n = -Z^2/2(n+\ell+1)^2. \tag{21}$$

On the other hand, the new problem whose phase shift is given by Eq. (18) has the following energy spectrum

$$E_n = \lambda^2/2\left[e^{-\beta\left(n+\ell+\lambda^{-1}\sqrt{V_0}\right)^2} - 1\right], \tag{22}$$

which does not belong to any of the known classes of energy spectra for exactly solvable systems. However, taking $\beta = (\lambda/Z)^2$ and $V_0 = \lambda^2$, then in the limit $\lambda \to 0$ this problem becomes the Coulomb problem. Another example in this class of new integrable systems is one that corresponds to $\theta = \pi/2$ and $y = (\lambda/k)^{2\ell+1}$, which has the following energy spectrum

$$E_n = -\tfrac{1}{2}\lambda^2(n+\mu)^{-1/(\ell+\tfrac{1}{2})}. \tag{23}$$

## 4.2 The continuous dual Hahn polynomial class:

Here, the energy polynomials are the orthonormal version of the continuous dual Hahn polynomials [21], which we write as

$$S_n^\mu(y^2;a,b) = \sqrt{\frac{(\mu+a)_n(\mu+b)_n}{n!(a+b)_n}}\,_3F_2\left({-n,\mu+iy,\mu-iy \atop \mu+a,\mu+b}\bigg|1\right), \tag{24}$$

where $(z)_n = z(z+1)(z+2)...(z+n-1) = \frac{\Gamma(n+z)}{\Gamma(z)}$, $y > 0$ and $\{\mu,a,b\}$ are positive except for a pair of complex conjugates with positive real parts. These satisfy the following symmetric three-term recursion relation

$$y^2 S_n^\mu = \left[(n+\mu+a)(n+\mu+b) + n(n+a+b-1) - \mu^2\right] S_n^\mu$$
$$- \sqrt{n(n+a+b-1)(n+\mu+a-1)(n+\mu+b-1)}\, S_{n-1}^\mu \tag{25}$$
$$- \sqrt{(n+1)(n+a+b)(n+\mu+a)(n+\mu+b)}\, S_{n+1}^\mu$$

The corresponding normalized weight function reads as follows

$$\rho^\mu(y;a,b) = \frac{1}{2\pi}\frac{|\Gamma(\mu+iy)\Gamma(a+iy)\Gamma(b+iy)/\Gamma(2iy)|^2}{\Gamma(\mu+a)\Gamma(\mu+b)\Gamma(a+b)}, \tag{26}$$

If we define $\tilde{S}_n^\mu(y^2;a,b) = \frac{(\mu+a)_n(\mu+b)_n}{n!(a+b)_n}\,_3F_2\left({-n,\mu+iy,\mu-iy \atop \mu+a,\mu+b}\bigg|1\right)$, then the generating function becomes

$$\sum_{n=0}^\infty \tilde{S}_n^\mu(y^2;a,b)t^n = (1-t)^{-\mu+iy}\,_2F_1\left({a+iy,b+iy \atop a+b}\bigg|t\right). \tag{27}$$

Again, we use the Darboux's method to obtain the asymptotic formula. This is carried out in the Appendix where the relevant result is given as Eq. (A16). Noting that $\ln n \approx o(n^\xi)$ for any $\xi > 0$, then we extract the following phase shift

$$\delta(\varepsilon) = \arg\{\Gamma(2iy)/\Gamma(\mu+iy)\Gamma(a+iy)\Gamma(b+iy)\}. \tag{28}$$

Now, if we choose $y = |k|/\alpha$, $a = \mu$ and $b = \tfrac{1}{2} - \alpha^{-1}\beta\sqrt{2V_0}$, where $\{\alpha,\beta,V_0\}$ are real positive parameters such that $b < 0$, then the scattering phase shift is

–9–

$$\delta(\varepsilon) = \arg\Gamma\left(2i\alpha^{-1}|k|\right) - 2\arg\Gamma\left(\mu + i\alpha^{-1}|k|\right)$$
$$-\arg\Gamma\left[\tfrac{1}{2} + \alpha^{-1}\left(i|k| - \beta\sqrt{2V_0}\right)\right] \tag{29}$$

whereas, bound states energies are obtained from (A18) as $y^2 = -(n+b)^2$ giving the following energy spectrum

$$E_n = -\frac{\alpha^2}{2}\left(n + \tfrac{1}{2} - \alpha^{-1}\beta\sqrt{2V_0}\right)^2, \tag{30}$$

and $n = 0,1,..,n_{max}$, where $n_{max}$ is the largest integer less than or equal to $-b$. This energy spectrum formula and the scattering phase shift (29) are associated with the 1D Morse potential $V(x) = V_0\left(e^{-2\alpha x} - 2\beta e^{-\alpha x}\right)$ for $b = -n_{max} - \mu$ and $1 > \mu \geq 0$. The corresponding $m^{\text{th}}$ bound states is obtained by substituting $i(m+b)$ for $y$ in the polynomial definition (24) giving

$$\psi_m(x) = A_m \sum_{n=0}^{N} S_n^\mu\left(-(m+b)^2;\mu,b\right)\phi_n^\lambda(x)$$
$$= A_m \sum_{n=0}^{N} \sqrt{\frac{(2\mu)_n}{n!}}\,{}_3F_2\left(\begin{matrix}-n,2\mu-m+N,m-N\\2\mu,-N\end{matrix}\bigg|1\right)\phi_n^\lambda(x) \tag{31}$$

where $N = n_{max}$ and the normalization factor $A_m = \sqrt{\rho_m^\mu}$. If we define the integer $\ell = N - m = N, N-1,...,0$ in the hypergeometric function in (31) then it becomes ${}_3F_2\left(\begin{matrix}-n,-\ell,\ell+2\mu\\2\mu,-N\end{matrix}\bigg|1\right)$, which is a special case of the discrete dual Hahn polynomial discussed below and we obtain $A_m^2 = \rho_m^\mu = 2(N!)\frac{(\ell+\mu)(2\mu)_\ell(N-\ell+1)_\ell}{(\ell+2\mu)_{N+1}(\ell!)^2}$. Moreover, we can show that $\langle\psi_\ell|\bar\psi_{\ell'}\rangle = \delta_{\ell,\ell'}$. If, on the other hand, our choice of the physical parameters is as follows: $y = \sqrt{\beta^{-1}\ln(1+k^2/\alpha^2)}$ and $\mu < 0$, then the associated energy spectrum is obtained by setting $iy = -(n+\mu)$ giving

$$E_n = \frac{\alpha^2}{2}\left[e^{-\beta(n+\mu)^2} - 1\right], \tag{32}$$

where $n = 0,1,..,N$, where $N$ is the largest integer less than or equal to $-\mu$. The corresponding $m^{\text{th}}$ bound state wavefunction is obtained by substituting $i(m+\mu)$ for $y$ in the polynomial definition (24) giving $S_n^\mu\left(-(m+\mu)^2;a,b\right) = \sqrt{\frac{(\mu+a)_n(\mu+b)_n}{n!(a+b)_n}} \times {}_3F_2\left(\begin{matrix}-n,-m,2\mu+m\\\mu+a,\mu+b\end{matrix}\bigg|1\right)$. This is proportional to the discrete dual Hahn polynomial, which we define here as $R_n^N(m;\alpha,\beta) = \sqrt{\frac{(\alpha+1)_n(\beta+1)_{N-n}}{n!(N-n)!}}\,{}_3F_2\left(\begin{matrix}-n,-m,m+\alpha+\beta+1\\\alpha+1,-N\end{matrix}\bigg|1\right)$, where $n,m = 0,1,2,..,N$, and either $\alpha,\beta > -1$ or $\alpha,\beta < -N$. These polynomials satisfy a discrete orthogonality relation that reads as follows (Eq. 1.6.1 and Eq. 1.6.2 in [21])

$$\sum_{m=0}^{N} \rho^N(m;\alpha,\beta)\, R_n^N(m;\alpha,\beta)\, R_{n'}^N(m;\alpha,\beta) = \delta_{n,n'}, \tag{33}$$

where $\rho^N(m;\alpha,\beta) = (N!)\frac{(2m+\alpha+\beta+1)(\alpha+1)_m(N-m+1)_m}{(m+\alpha+\beta+1)_{N+1}(\beta+1)_m m!}$. The polynomial parameters are related to the physical parameters as follows: $\alpha = \mu + a - 1$, $\beta = \mu - a$ and $N = $ $-\mu-b$; thus, it is required that $1 > b \geq 0$ and $1-b < a < 1+b$. Finally, the $m^{\text{th}}$ bound state wavefunction becomes

$$\psi_m(x) = \sqrt{\rho^N(m;\alpha,\beta)} \sum_{n=0}^{N} R_n^N(m;\alpha,\beta) \phi_n^\lambda(x). \tag{34}$$

Using $\langle \phi_n^\lambda | \bar{\phi}_{n'}^\lambda \rangle = \delta_{n,n'}$ and the *dual* orthogonality of $R_n^N(m;\alpha,\beta)$,[7]

$$\sum_{n=0}^{N} R_n^N(m;\alpha,\beta) R_n^N(m';\alpha,\beta) = \frac{\delta_{m,m'}}{\rho^N(m;\alpha,\beta)}, \tag{35}$$

one can proof that these bound states are orthonormal; that is, $\langle \psi_m | \bar{\psi}_{m'} \rangle = \delta_{m,m'}$. Now, the system with energy spectrum formula (32) and corresponding bound state wave function (34) is not an element in the known class of exactly solvable problems. Another such example with $\mu < 0$ but with $y = (|k|/\alpha)^{2\ell+1}$ has the following finite energy spectrum

$$E_n = -\frac{1}{2}\alpha^2(n+\mu)^{1/(\ell+\frac{1}{2})}. \tag{36}$$

where $n = 0,1,..,N$. In fact, with $y = |k|/\alpha$ and $\mu \neq a \neq b \neq \mu$ formula (A18) could result in a more general 3-fold energy spectrum consisting of the union of three sets of spectral lines as follows

$$E_{\ell,m,n} = -\frac{\alpha^2}{2}\left[\{(\ell+\mu)^2\} \cup \{(m+a)^2\} \cup \{(n+b)^2\}\right], \tag{37}$$

where $\{\ell,m,n\} = 0,1,2,..,N_{\{\ell,m,n\}}$. Finally, if the physical parameters are such that $\mu < 0$ and $a+\mu$, $b+\mu$ are positive or a pair of complex conjugates with positive real parts, then the system will have continuum scattering states as well as a finite number of discrete bound states and the corresponding energy polynomials satisfy the following generalized orthogonality relation (Eq. 1.3.3 in [21])

$$\int_0^\infty \rho^\mu(y;a,b) S_n^\mu(y^2;a,b) S_{n'}^\mu(y^2;a,b) dy - 2\frac{\Gamma(a-\mu)\Gamma(b-\mu)}{\Gamma(a+b)\Gamma(1-2\mu)} \times$$

$$\sum_{m=0}^{N} (-1)^m (m+\mu) \frac{(\mu+a)_m (\mu+b)_m (2\mu)_m}{(\mu-a+1)_m (\mu-b+1)_m m!} S_n^\mu(m;a,b) S_{n'}^\mu(m;a,b) = \delta_{n,n'} \tag{38}$$

where $S_n^\mu(m;a,b) \equiv S_n^\mu(-(m+\mu)^2;a,b)$.

## 5. SYSTEMS WITH DISCRETE ENERGY SPECTRA ASSOCIATED WITH POTENTIAL FUNCTIONS

The wavefunction for such systems could be written as an element in a discrete set of states $\{\psi_n(x)\}_{n=0}^\infty$. Each one of these discrete elements is written in terms of orthogonal polynomials in configuration space (not in energy space) that satisfy a linear second order differential equation. These are the classical orthogonal polynomials $\{P_n(y)\}_{n=0}^\infty$, which are associated with positive definite measures (weight/density functions) $\omega(y)$

---
[7] See un-numbered equation at the bottom of page 36 in [21]

that are continuous over a connected interval $[y_-, y_+]$ such that (see, for example [12,13])

$$\int_{y_-}^{y_+} \omega(y) P_n(y) P_m(y) dy = \zeta_n \delta_{nm}, \qquad (39)$$

where $\zeta_n > 0$. Now, let the variable $y$ (the polynomial argument) be related to the configuration space coordinate $x$ as $y = y(x)$, then the orthogonality of the bound state wavefunction, $\lambda \int_{x_-}^{x_+} \psi_n(x) \psi_m(x) dx = \delta_{nm}$, together with Eq. (39) imply that we can take the representation $\psi_n(x) = \sqrt{y'\omega(y)/\lambda\zeta_n}\, P_n(y)$. Note that since $\omega(y_\pm) = 0$, then the discrete bound states satisfy the boundary condition $\psi_n(x_\pm) = 0$ provided that $y'(x_\pm)$ are finite. To complete the specification of the physical system, we need to provide only the set of energy eigenvalues $\{E_n\}_{n=0}^{\infty}$. We could obtain those by utilizing the differential equation satisfied by the orthogonal polynomials $\{P_n(y)\}_{n=0}^{\infty}$ and comparing it to the wave equation for $\psi_n(x)$, $\left[-\frac{1}{2}\frac{d^2}{dx^2} + V(x) - E_n\right]\psi_n(x) = 0$. Replacing $\psi_n(x)$ by $\sqrt{y'\omega(y)/\lambda\zeta_n}\, P_n(y)$, this wave equation becomes an equation in terms of the variable $y$ for the polynomials $P_n(y)$ as follows

$$\left\{ y'^2 \frac{d^2}{dy^2} + \left(2y'' + y'^2 \frac{\omega'}{\omega}\right)\frac{d}{dy} + y'^2\left[\frac{1}{4}\left(\frac{\omega'}{\omega}\right)^2 + \frac{1}{2}\left(\frac{\omega'}{\omega}\right)'\right] \right.$$

$$\left. + y''\left(\frac{\omega'}{\omega}\right) + \frac{1}{4}\left(\frac{y''}{y'}\right)^2 + \frac{1}{2}\left(\frac{y''}{y'}\right)' - 2V(x) + 2E_n \right\} P_n(y) = 0 \qquad (40)$$

where the prime denotes the derivative with respect to the argument (i.e., $y' = \frac{dy}{dx}$ and $\omega' = \frac{d\omega}{dy}$).

**Examples**:
As an example, let us consider the Hermite polynomials $H_n(y)$ defined on the whole real line and take $y(x) = \kappa x$. The orthogonality of these polynomials reads as follows (see page 6 of [12])

$$\int_{-\infty}^{+\infty} e^{-y^2} H_n(y) H_m(y) dy = \sqrt{\pi}\, 2^n \Gamma(n+1) \delta_{nm}. \qquad (41)$$

Thus, $\omega(y) = e^{-y^2}$ and we can write $\psi_n(x) = \frac{\pi^{-1/4}}{\sqrt{2^n \Gamma(n+1)}} e^{-\kappa^2 x^2/2} H_n(\kappa x)$. Equation (40) becomes

$$\left[\frac{d^2}{dy^2} - 2y\frac{d}{dy} + y^2 - 1 + \frac{2}{\kappa^2}(E_n - V(x))\right] H_n(y) = 0. \qquad (42)$$

Comparing this with the differential equation for the Hermite polynomials, $H_n(y)'' - 2y H_n(y)' + 2n H_n(y) = 0$, we conclude that $V(x) = \frac{1}{2}\kappa^4 x^2$ and $E_n = \kappa^2\left(n + \frac{1}{2}\right)$. Therefore, this corresponds to the one-dimensional harmonic oscillator problem.

As a second example, we consider the ultra-spherical (Gegenbauer) polynomials $C_n^\mu(y)$ where $\mu > -\frac{1}{2}$, $y \in [-1,+1]$. The corresponding orthogonality relation is (see pages 143-153 of [12])

$$\int_{-1}^{+1}(1-y^2)^{\mu-\frac{1}{2}} C_n^\mu(y) C_m^\mu(y) dy = \frac{\pi/2^{2\mu-1}}{\Gamma(\mu)^2} \frac{\Gamma(n+2\mu)}{(n+\mu)\Gamma(n+1)} \delta_{nm}, \qquad (43)$$

whereas its differential equation reads as follows

$$\left[(1-y^2)\frac{d^2}{dy^2} - (2\mu+1)y\frac{d}{dy} + n(n+2\mu)\right] C_n^\mu(y) = 0. \qquad (44)$$

Thus, the weight function is $\omega^\mu(y) = (1-y^2)^{\mu-\frac{1}{2}}$ and comparing Eq. (44) with Eq. (40) we must choose $y'^2 \propto 1-y^2$ whose solution could be taken as $y = \sin(\pi x/a)$, where $x \in \left[-\frac{a}{2}, +\frac{a}{2}\right]$. Consequently, Eq. (40) becomes as follows

$$\left\{(1-y^2)\frac{d^2}{dy^2} - (2\mu+1)y\frac{d}{dy} + \frac{\mu(\mu-1)}{1-y^2}\right.$$
$$\left. - \mu^2 + 2(a/\pi)^2 [E_n - V(x)]\right\} C_n^\mu(y) = 0 \qquad (45)$$

Comparing this with Eq. (44), we conclude that $V(x) = \frac{1}{2}\left(\frac{\pi}{a}\right)^2 \frac{\mu(\mu-1)}{\cos^2(\pi x/a)}$ and $E_n = \frac{1}{2}\left(\frac{\pi}{a}\right)^2 (n+\mu)^2$, which correspond to the Pöschl-Teller potential. Moreover, the orthogonal discrete bound states functions become

$$\psi_n(x) = 2^\mu \Gamma(\mu) \sqrt{\frac{(n+\mu)\Gamma(n+1)}{2\pi \Gamma(n+2\mu)}} \left[\cos(\pi x/a)\right]^\mu C_n^\mu(\sin(\pi x/a)). \qquad (46)$$

One of the exceptional examples that belong not only to the class of problems discussed above in Section 4 but also to the present class is the Coulomb problem. This one is associated with the Laguerre polynomials $L_n^\nu(y)$ defined on the semi-infinite real line $y \geq 0$ and we take $y(x) = \mu x$, where $\mu > 0$ and $\nu > -1$. Orthogonality of these polynomials read (see page 44 of [12])

$$\int_0^\infty y^\nu e^{-y} L_n^\nu(y) L_m^\nu(y) dy = \frac{\Gamma(n+\nu+1)}{\Gamma(n+1)} \delta_{nm}. \qquad (47)$$

Thus, $\omega(y) = y^\nu e^{-y}$ and we write $\psi_n(x) = \sqrt{\frac{\Gamma(n+1)}{\Gamma(n+\nu+1)}} (\mu x)^{\frac{\nu}{2}} e^{-\mu x/2} L_n^\nu(\mu x)$. Equation (40) becomes

$$\left[\frac{d^2}{dy^2} + \left(\frac{\nu}{y}-1\right)\frac{d}{dy} + \frac{\nu(\nu-2)}{4y^2} - \frac{\nu}{2y} + \frac{1}{4} + \frac{2}{\mu^2}(E_n - V(x))\right] L_n^\nu(y) = 0. \qquad (48)$$

Comparing this with the differential equation of the Laguerre polynomials, $\left[y\frac{d^2}{dy^2} + (\nu+1-y)\frac{d}{dy} + n\right] L_n^\nu(y) = 0$, we see that our definition of the basis must be modified to read $\psi_n(x) = \sqrt{\frac{\Gamma(n+1)}{\Gamma(n+\nu+1)}} (\mu x)^{\frac{\nu+1}{2}} e^{-\mu x/2} L_n^\nu(\mu x)$ in which case the set $\{\psi_n(x)\}_{n=0}^\infty$

–13–

becomes tri-thogonal not orthogonal. That is, $\langle\psi_n|\psi_m\rangle = \Lambda_{nm}$, where $\Lambda$ is the matrix representation of the identity, which is tridiagonal and symmetric as follows

$$\Lambda_{nm} = (2n+\nu+1)\delta_{n,m} - \sqrt{n(n+\nu)}\delta_{n,m+1} - \sqrt{(n+1)(n+\nu+1)}\delta_{n,m-1}. \qquad (49)$$

Moreover, the wave equation becomes

$$\left[\frac{d^2}{dy^2} + \left(\frac{\nu+1}{y}-1\right)\frac{d}{dy} + \frac{\nu^2-1}{4y^2} - \frac{\nu+1}{2y} + \frac{1}{4} + \frac{2}{\mu^2}(E_n - V(x))\right]L_n^\nu(y) = 0, \qquad (50)$$

giving $V(x) = \frac{\ell(\ell+1)}{2x^2} + \frac{Z}{x}$, where $\ell$ is the angular momentum quantum number. Thus, we obtain $\nu = 2\ell+1$, $\mu = -2Z/(n+\ell+1)$ and $E_n = -\frac{1}{2}[Z/(n+\ell+1)]^2$, which correspond to the radial 3D Coulomb problem for an electric charge $Z < 0$.

Another problem in this class, which is also associated with the Laguerre polynomial $L_n^\nu(y)$ with $\nu > 0$, is the 1D Morse potential $V(x) = V_0(e^{-2\alpha x} - 2\beta e^{-\alpha x})$ with $y(x) = \mu e^{-\alpha x}$. Following the same procedure as in the Coulomb problem, we obtain the energy spectrum formula (30) with $\mu = 2\alpha^{-1}\sqrt{2V_0}$ and $\nu = \alpha^{-1}\beta\sqrt{2V_0} - n - \frac{1}{2}$. Since $\nu > 0$, then $n < \alpha^{-1}\beta\sqrt{2V_0} - \frac{1}{2}$. The corresponding bound state wavefunction is

$$\psi_n(x) = \sqrt{\frac{\Gamma(n+1)}{\Gamma(n+\nu_n+1)}}\, y^{\frac{\nu_n-1}{2}} e^{-y/2} L_n^{\nu_n}(y). \qquad (51)$$

Table 1 shows additional examples in this class of systems whose potential functions can be realized analytically that are also associated with the classical orthogonal polynomials.

## 6. CONCLUSION AND DISCUSSION

In this work, we presented an alternative formulation of quantum mechanics in which the potential function is not required. The aim is to obtain a set of analytically realizable systems, which is larger than that in the standard formulation. We start by proposing a wavefunction, which is constructed as an infinite bounded sum of polynomials that are orthogonal in the energy space. Using the asymptotics of these polynomials, we obtain the scattering phase shift, bound states and resonances. Thus, instead of proposing a potential function to model the physical system we propose a set of orthogonal polynomials in the energy as expansion coefficients of the wavefunction. In addition to the well-known classes of exactly solvable quantum systems, this formulation enables one to present new solvable systems. Such systems are not necessarily associated with any given realization of analytic potential functions. Two set of examples were given to illustrate the utility and viability of this formalism. Finally, we like to add the following relevant remarks and points of discussion:

(1) We believe that this formulation could easily be extended to relativistic quantum mechanics. Our recent work on the Dirac-Coulomb problem suggests that this is feasible [23].

(2) One may also attempt to construct a quantum field theory based on this formalism in which the polynomials of the first and second kind, $P_n^\mu(\varepsilon)$ and $Q_n^\mu(\varepsilon)$, act on the $n^{\text{th}}$ vacuum (kinematic) state $|\phi_n^\lambda\rangle$ as creation and annihilation operators of physical quanta with properties $\{\mu\}$. For that, one needs to define a pairing operator $[\ ,\ ]$ such that $\left[P_n^\mu(\varepsilon), Q_n^\mu(\varepsilon)\right] \sim 1$. As an example, one may take the odd (even) Wronskian, $P_n^\mu Q_{n-1}^\mu \pm Q_n^\mu P_{n-1}^\mu$, as representation of this pairing operator for fermion (boson) fields where for the odd fermions $Q_n^\mu \to (-1)^n Q_n^\mu$.

(3) In the examples of section 4, we made a seemingly arbitrary choice of energy polynomials and only after making a particular choice of parameters that we were able to obtain the properties of the corresponding physical system. However, the physical insight one develops when dealing with the potential function in standard quantum mechanics such as the properties of bound states and resonances, tunneling effects, location of stable points, phase transitions, classical turning points etc., is to be replaced in our approach with another that links the properties of orthogonal polynomials to these physical phenomena. Such properties include, but not limited to, the shape of weight functions, nature of the generating functions, distribution and density of the zeros, recursion relations, asymptotics, differential or difference equations, etc. In addition to the requirement that the energy polynomials of physical relevance have the asymptotic behavior (9a), two other associated polynomials must also coexist. The first one is the polynomial of the second kind with the asymptotic behavior shown as (9b). The second one is the discrete version of the polynomial, which is appropriate for the description of bound states.

(4) The wave equation in the standard formulation is replaced by the three-term recursion relation of the orthogonal polynomials $\{P_n^\mu(\varepsilon), Q_n^\mu(\varepsilon)\}$. To show that, let us assume that the action of the Hamiltonian operator on the basis elements is as follows
$$H|\phi_n^\lambda\rangle = a_n|\phi_n^\lambda\rangle + b_{n-1}|\phi_{n-1}^\lambda\rangle + b_n|\phi_{n+1}^\lambda\rangle, \qquad (52)$$
where the constants $\{a_n, b_n\}$ depend on the physical parameters $\{\mu\}$ contained in $H$. If $H$ is written in the conventional format as the sum of the kinetic energy operator and the potential function, and $|\phi_n^\lambda\rangle$ is one of the basis functions given in terms of the classical orthogonal polynomials as shown at the beginning of Section 2 above, then Eq. (52) is true only for a small set of potentials (see Ref. [24] for illustrations). Nonetheless, acting by $H$ on the wavefunction expansion $|\psi(x,\varepsilon)\rangle = \sqrt{\rho^\mu(\varepsilon)} \sum_{n=0}^\infty P_n^\mu(\varepsilon)|\phi_n^\lambda(x)\rangle$ and using (52) together with the wave equation $H|\psi\rangle = E|\psi\rangle$, we obtain
$$E\, P_n^\mu(\varepsilon) = a_n P_n^\mu(\varepsilon) + b_{n-1} P_{n-1}^\mu(\varepsilon) + b_n P_{n+1}^\mu(\varepsilon), \qquad (53)$$
where $E = E(\mu, \varepsilon)$. Therefore, the wave equation becomes equivalent to the three-term recursion relation satisfied by the energy polynomials $\{P_n^\mu(\varepsilon), Q_n^\mu(\varepsilon)\}$.

(5) In all the examples of Section 4 above, no resonances were found. This is so because all zeros of the scattering amplitudes $A(\varepsilon)$ obtained from the asymptotics of the energy polynomials were real. Here, we give an example where resonances occur and consider the case of the Meixner-Pollaczek polynomial $P_n^\mu(y;\theta)$. If we choose the physical parameters as $\mu = \alpha/\sqrt{2E}$, $y = \beta/\alpha$ and $\theta = \pi/2$, where $\alpha > \beta > 0$, then the scattering phase shift becomes $\delta(\varepsilon) = \arg\Gamma(\alpha/k + i\beta/\alpha)$, whereas bound states and resonances occur for $\mu + iy = -n$ giving

$$E_n = \frac{\frac{1}{2}(n^2\alpha^2 - \beta^2) - in\alpha\beta}{\left(n^2 + \beta^2/\alpha^2\right)^2}. \tag{54}$$

The result is a single bound state at $E_0 = -\frac{1}{2}(\alpha^2/\beta)^2$ and an infinite number of resonances at $E = E_1, E_2, \ldots$. For a given choice of values of the parameters $\alpha$ and $\beta$, Figure 1 shows the bound state and resonance energies in the complex energy plane. The $m^{th}$ resonance wavefunction is obtained by substituting $\theta = \pi/2$ and $\mu = \alpha/\sqrt{2E_m}$ in Eq. (20). It is a complex wavefunction except for $m = 0$ where it is real and with $\mu = \alpha/\sqrt{-2E_m}$. For large $x$, it becomes vanishingly small. Figure 2 shows $|\psi_m(x)|$ for the bound state ($m = 0$) as well as few of the lowest order resonances. For resonances, the figure shows only the initial states at $t = 0$. An interesting situation occurs if $\beta > \alpha > 0$. In this case, some of the lowest order resonances will be located in the third quadrant of the complex energy plane. That is, both the real and imaginary parts of the resonance energies are negative; a situation that we may refer to as "bound-state-embedded resonances" (see Fig. 3)

(6) We plan to follow this work with another where we show how, if possible, to reconstruct (analytically or numerically) the potential function so that a correspondence with the standard formulation of quantum mechanics could be established. Such work is very appealing, especially for systems with new scattering phase shifts and/or energy spectra like those found in Section 4 above. However, we may discover that forcing such correspondence may lead to energy dependent potentials, nonlocal potentials, or to differential wave equations that are higher than second order (possibly of an infinite order). An outline of one such investigation for systems with short-range potentials goes as follows: The left-hand side of Eq. (52) could written as $H|\phi_n^\lambda\rangle = T|\phi_n^\lambda\rangle + V|\phi_n^\lambda\rangle$, where $T$ is the kinetic energy operator. Projecting from left by $\langle\bar{\phi}_m^\lambda|$, we obtain

$$V_{mn} = \langle\bar{\phi}_m^\lambda|V|\phi_n^\lambda\rangle = \langle\bar{\phi}_m^\lambda|H|\phi_n^\lambda\rangle - \langle\bar{\phi}_m^\lambda|T|\phi_n^\lambda\rangle. \tag{55}$$

From Eq. (52), we can write $\langle\bar{\phi}_m^\lambda|H|\phi_n^\lambda\rangle = a_n\delta_{m,n} + b_{n-1}\delta_{m,n-1} + b_n\delta_{m,n+1}$. Since the action of the space derivative in $T$ (or more generally, the shift operators $\pm\frac{d}{dx} + \frac{\ell}{x}$) on the $x$-dependent classical polynomials in $\phi_n^\lambda$ is known or could be derived, then the matrix elements of the kinetic energy operator can be obtained. Therefore, the matrix elements of the potential become readily available. Now, with knowledge of: (1) a complete $L^2$ basis functions, and (2) the matrix elements of the potential in this basis, one should, in principle, be able to construct (numerically or analytically)

the potential function. For example, in such an approach one obtains the following local approximation for the potential function

$$V(x) \cong \sum_{n,m=0}^{N} V_{nm}\, \phi_n^\lambda(x)\bar{\phi}_m^\lambda(x) \Big/ \sum_{n=0}^{N} \phi_n^\lambda(x)\bar{\phi}_n^\lambda(x) \qquad (56)$$

where $N$ is some large enough integer.

(7) Perturbative and non-perturbative calculations in this formalism could be developed by writing the total Hamiltonian as $H = H_0 + \eta W$, where $H_0$ is the reference Hamiltonian whose action on the basis elements, which is given by Eq. (52), results in one of the known recursion relations whose exact polynomial solution is given. On the other hand, $W$ is an additional interaction Hamiltonian whose action on the basis is the same as (52) but with new recursion coefficients $\{\alpha_n, \beta_n\}$ making the total wave equation equivalent to the following recursion relation that replaces (53)

$$E\, P_n^\mu = (a_n + \eta\alpha_n)P_n^\mu + (b_{n-1} + \eta\beta_{n-1})P_{n-1}^\mu + (b_n + \eta\beta_n)P_{n+1}^\mu. \qquad (57)$$

For perturbative calculation $\eta \ll 1$, whereas for non-perturbative calculation this restriction is not needed but replaced by the requirement that $W$ is short-range, which means that either $(\alpha_n, \beta_n) \to 0$ fast enough as $n \to \infty$ or $(\alpha_n, \beta_n) = 0$ for $n$ larger than a given finite integer. Now, a numerical routine could be developed to solve (57) that build on knowledge of the exact solution of (53). One such routine, which is based on the J-matrix method of scattering, is outlined in [7]. It is worth noting that the general action of the interaction Hamiltonian $W$ on the basis does not have to be represented by a tridiagonal matrix with the elements $\{\alpha_n, \beta_n\}$ as shown above in Eq. (57) but can be any finite Hermitian matrix, which could easily be handled by the numerical routine.

(8) An additional third class of exactly solvable problems in our formulation of quantum mechanics is a 4-parameter class associated with the Wilson polynomial, which is written in terms of the hypergeometric function ${}_4F_3$ as follows

$$W_n^\mu(y^2; v, a, b) = \frac{(\mu+a)_n(\mu+b)_n}{(a+b)_n\, n!}\, {}_4F_3\!\left(\begin{array}{c}-n, n+\mu+v+a+b, \mu+iy, \mu-iy\\ \mu+v, \mu+a, \mu+b\end{array}\Big| 1\right), \qquad (58)$$

where $\mathrm{Re}(\mu, v, a, b) > 0$ and non-real parameters occur in conjugate pairs. From the asymptotic formula of this polynomial, one concludes that bound states could exist if $iy = -(n+\mu)$. These are finite in number and written in terms of the discrete Racah polynomial, which could be written as follows:

$$R_n^N(m; \alpha, \beta, \gamma) = \frac{(-N)_n(\beta+1)_n}{(-N-\gamma)_n\, n!}\, {}_4F_3\!\left(\begin{array}{c}-n, -m, n+\alpha-\gamma-N, m+\beta+\gamma\\ \alpha+1, \beta+1, -N\end{array}\Big| 1\right), \qquad (59)$$

where $n, m = 0, 1, 2, .., N$. Properties of these polynomials are given in §1.1 and §1.2 of [21], respectively. Due to space limitation, we leave the details of the physical systems associated with this very interesting class for a future study.

(9) The logarithmic dependence in the argument of the cosine in the asymptotic formula (A4) and (A16) is very interesting and a source of curiosity. Despite overlooking the full implication of this behavior through our use of the asymptotic relation $\ln n \approx o(n^\xi)$ for any $\xi > 0$, the understanding of the source of this logarithmic behavior and its physical implication must be very fruitful.

(10) The bound states energy spectra that we have, so far, been able to find are all functions of $(n+\alpha)^{\pm 2}$, where $\alpha$ is some real parameter. This is because all polynomials that we looked at have asymptotics with amplitudes that contain the factor $|\Gamma(\alpha+ix)|^{-1}$. The Wilson polynomials will also have the same type of spectrum. On the other hand, there are other types of energy spectra such as the one associated with the Hulthén and Rosen-Morse potentials. This is of the form $E_n \sim \left(n+\alpha+\frac{\gamma}{n+\alpha}\right)^2$, where $\gamma$ is another real parameter[8]. We only know the recursion relation of the corresponding energy polynomials. However, their weight functions, generating functions, asymptotic properties, etc. are yet to be derived. The recursion relations for these polynomials are related to that of the Jacobi polynomial, $P_n^{(\mu,\nu)}(y)$, which is written as

$$y P_n^{(\mu,\nu)}(x) = A_n P_n^{(\mu,\nu)}(y) + B_{n-1} P_{n-1}^{(\mu,\nu)}(y) + C_n P_{n+1}^{(\mu,\nu)}(y), \tag{60}$$

where,

$$A_n = \frac{\nu^2-\mu^2}{(2n+\mu+\nu)(2n+\mu+\nu+2)}, \tag{61a}$$

$$B_n = \frac{2(n+\mu+1)(n+\nu+1)}{(2n+\mu+\nu+2)(2n+\mu+\nu+3)}, \tag{61b}$$

$$C_n = \frac{2(n+1)(n+\mu+\nu+1)}{(2n+\mu+\nu+1)(2n+\mu+\nu+2)}. \tag{61c}$$

Now, define the polynomial $Q_n^{(\mu,\nu)}(y;\lambda)$ by the following recursion relation (see Eq. (6.11) in [24] and Eq. (B.6) in [25])

$$\left(y-\lambda D_n^2\right)Q_n^{(\mu,\nu)}(y;\lambda) = A_n Q_n^{(\mu,\nu)}(y;\lambda) + B_{n-1} Q_{n-1}^{(\mu,\nu)}(y;\lambda) + C_n Q_{n+1}^{(\mu,\nu)}(y;\lambda) \tag{62}$$

for $n=1,2,..$ and with $Q_0^{(\mu,\nu)}(y)=1$, $Q_1^{(\mu,\nu)}(y)=\left(y-\lambda D_0^2 - A_0\right)/C_0$, where $D_n = n+\frac{\mu+\nu+1}{2}$. The spectrum associated with this polynomial is finite and of the form $E_n \sim \left(n+\alpha+\frac{\gamma}{n+\alpha}\right)^2$ at which some of the parameters $\mu$ and/or $\nu$ become $n$-dependent (see Eq. (6.13) in [24]). Another polynomial example with the same type of spectrum is $H_n^{(\mu,\nu)}(y;\lambda)$, which is defined by (see Eq. (6.17) in [24] but replace $y$ by $\frac{1+y}{2}$):

$$\left(y+F_n\right)H_n^{(\mu,\nu)}(y;\lambda) = G_n A_n H_n^{(\mu,\nu)}(y;\lambda) + G_{n-1} B_{n-1} H_{n-1}^{(\mu,\nu)}(y;\lambda) + G_n C_n H_{n+1}^{(\mu,\nu)}(y;\lambda) \tag{63}$$

where $F_n = \frac{n(n+\mu)}{2n+\mu+\nu}$ and $G_n = \lambda + \left(n+1+\frac{\mu+\nu}{2}\right)^2 = \lambda + \left(D_n+\frac{1}{2}\right)^2$.

**Acknowledgments:**
The support provided by the Saudi Center for Theoretical Physics (SCTP) is highly appreciated by the two Authors. Ismail acknowledges partial support from the DSFP program at King Saud University. We are also grateful to Andrey M. Shirokov (Moscow State University) and to Alexander E. Kaplan (Johns Hopkins University) for fruitful communications.

---

[8] This spectrum is the sum of $(n+\alpha)^2$ and $(n+\alpha)^{-2}$; nonetheless, the relation of the two systems with these spectra to the one with their sum is highly nontrivial.

## APPENDIX: ASYMPTOTICS OF ORTHOGONAL POLYNOMIALS

In this Appendix we give the details for deriving the asymptotic ($n \to \infty$) formulas for the Meixner-Pollaczek polynomials and the continuous dual Hahn polynomials. For that, we use Darboux's method (see Chapter 8 of [22]).

**The Meixner-Pollaczek polynomials**:
With the generating function (15), the dominant term in a comparison function is

$$\left(1-e^{2i\theta}\right)^{-\mu+iy}\left(1-te^{-i\theta}\right)^{-\mu-iy} + \left(1-e^{-2i\theta}\right)^{-\mu-iy}\left(1-te^{i\theta}\right)^{-\mu+iy}. \quad (A1)$$

Therefore, the asymptotics ($n \to \infty$) of $\tilde{P}_n^\mu(y;\theta)$ is

$$\tilde{P}_n^\mu(y;\theta) \approx \left(1-e^{2i\theta}\right)^{-\mu+iy}\frac{(\mu+iy)_n}{n!}e^{-in\theta} + \left(1-e^{-2i\theta}\right)^{-\mu-iy}\frac{(\mu-iy)_n}{n!}e^{in\theta} \quad (A2)$$

Using $\frac{\Gamma(n+a)}{\Gamma(n+b)} \approx n^{a-b}$, $(z)_n = \frac{\Gamma(n+z)}{\Gamma(z)}$, $a^{ib} = e^{ib\ln a}$, $\Gamma(z) = |\Gamma(z)|e^{i\arg\Gamma(z)}$, this formula becomes

$$\tilde{P}_n^\mu(y;\theta) \approx \frac{n^{\mu-1}}{|\Gamma(\mu+iy)|}\left[e^{iy\ln n}\,e^{-i\gamma}\,e^{-in\theta}\left(1-e^{2i\theta}\right)^{-\mu+iy} + \text{complex conjugate}\right]$$

$$= \frac{2n^{\mu-1}e^{(\pi/2-\theta)y}}{(2\sin\theta)^\mu|\Gamma(\mu+iy)|}\cos\left[n\theta + \gamma + \mu(\theta - \pi/2) - y\ln(2n\sin\theta)\right] \quad (A3)$$

where $\gamma = \arg\Gamma(\mu+iy)$. Thus, with $P_n^\mu(y;\theta) = \sqrt{\frac{\Gamma(2\mu)\Gamma(n+1)}{\Gamma(n+2\mu)}}\tilde{P}_n^\mu(y;\theta)$ we obtain

$$P_n^\mu(y;\theta) \approx \frac{2n^{-1/2}e^{(\pi/2-\theta)y}}{(2\sin\theta)^\mu|\Gamma(\mu+iy)|}\cos\left[n\theta + \gamma + \mu(\theta-\pi/2) - y\ln(2n\sin\theta)\right]. \quad (A4)$$

Therefore, the asymptotic relation for the continuum scattering states is as follows

$$\sqrt{\rho^\mu(y;\theta)}\,P_n^\mu(y;\theta) \approx \sqrt{\frac{2}{n\pi}}\cos\left[n\theta + \gamma + \mu(\theta-\pi/2) - y\ln(2n\sin\theta)\right]. \quad (A5)$$

The asymptotics of the more general Wilson polynomials were derived in [26]. Wilson's result however is not applicable because the Meixner-Pollaczek polynomials are a limiting case of the Wilson polynomials and we cannot simply interchange this limit with the asymptotic limit. Fields and Luke [27] also studied asymptotics of the more general hypergeometric function ${}_pF_q$ and it may be possible to extract (A5) from their work but this involved a complicated calculation. The relevant results of Fields and Luke are reproduced in §7.4.6 of [28]. It must be noted that Wilson's results do not follow from the Fields and Luke results.

Since bound states correspond to asymptotically zero norm states, then such condition is met if the amplitude of the asymptotic oscillations of the Meixner-Pollaczek polynomials given by (A4) vanishes. This occurs at the poles of the gamma function $\Gamma(\mu+iy)$ (i.e., if $\mu+iy$ becomes a negative integer or zero). That is, bound states and resonances correspond to the condition

$$\mu + iy = -n, \quad (A6)$$

–19–

where $n = 0, 1, 2, \ldots$, giving the bound states energy spectrum formula $y^2 = -(n+\mu)^2$.

**The continuous dual Hahn polynomials**:
To derive the large $n$ asymptotic formula we use again Darboux's method (Chapter 8 of [22]). We also need the contiguous relation (Eq. 7 of Ex. 21 in [29])

$$(a+b-c)\,_2F_1\left(\genfrac{}{}{0pt}{}{a,b}{c}\bigg|z\right) = a(1-z)\,_2F_1\left(\genfrac{}{}{0pt}{}{a+1,b}{c}\bigg|z\right) - (c-b)\,_2F_1\left(\genfrac{}{}{0pt}{}{a,b-1}{c}\bigg|z\right), \quad (A7)$$

and the Euler transformation [29]

$$_2F_1\left(\genfrac{}{}{0pt}{}{a,b}{c}\bigg|z\right) = (1-z)^{c-a-b}\,_2F_1\left(\genfrac{}{}{0pt}{}{c-a,c-b}{c}\bigg|z\right). \quad (A8)$$

Moreover, we will apply the Gauss sum (Theorem 18 in §32 of [29])

$$_2F_1\left(\genfrac{}{}{0pt}{}{a,b}{c}\bigg|1\right) = \frac{\Gamma(c)\Gamma(c-a-b)}{\Gamma(c-a)\Gamma(c-b)}, \quad \mathrm{Re}(c-a-b) > 0. \quad (A9)$$

Therefore, the right-hand side of the generating function (27) becomes

$$\frac{(1-t)^{-\mu+iy}}{2iy}\left\{(a+iy)(1-t)\,_2F_1\left(\genfrac{}{}{0pt}{}{a+1+iy,b+iy}{a+b}\bigg|t\right) - (a-iy)\,_2F_1\left(\genfrac{}{}{0pt}{}{a+iy,b-1+iy}{a+b}\bigg|t\right)\right\}. \quad (A10)$$

We use (A8) to write the first term in the curled bracket as

$$(a+iy)(1-t)^{-2iy}\,_2F_1\left(\genfrac{}{}{0pt}{}{b-1-iy,a-iy}{a+b}\bigg|t\right) \quad (A11)$$

This shows that the dominant term in a comparison function is

$$\frac{1}{2iy}\left\{(a+iy)(1-t)^{-\mu-iy}\,_2F_1\left(\genfrac{}{}{0pt}{}{a-iy,b-1-iy}{a+b}\bigg|t\right) - \text{with } y \to -y\right\} \quad (A12)$$

The Gauss sum (A9) evaluates the above $_2F_1$'s at $t = 1$. Therefore, we obtain

$$\tilde{S}_n^\mu(y^2;a,b) \approx \frac{1}{2iy}\left\{(a+iy)\frac{(\mu+iy)_n}{\Gamma(n+1)}\frac{\Gamma(a+b)\Gamma(1+2iy)}{\Gamma(a+1+iy)\Gamma(b+iy)} - \text{complex conjugate}\right\}$$

$$\approx \frac{\Gamma(a+b)\Gamma(2iy)n^{\mu+iy-1}}{\Gamma(a+iy)\Gamma(b+iy)\Gamma(\mu+iy)} + \text{complex conjugate} \quad (A13)$$

Let,

$$\Gamma(\mu+iy)\Gamma(a+iy)\Gamma(b+iy)/\Gamma(2iy) = |\Gamma(\mu+iy)\Gamma(a+iy)\Gamma(b+iy)/\Gamma(2iy)|e^{i\gamma} \quad (A14)$$

Therefore,

$$\tilde{S}_n^\mu(y^2;a,b) \approx \frac{2\Gamma(a+b)|\Gamma(2iy)|n^{\mu-1}}{|\Gamma(a+iy)\Gamma(b+iy)\Gamma(\mu+iy)|}\cos(y\ln n - \gamma) \quad (A15)$$

Noting that $S_n^\mu(y^2;a,b) = \sqrt{\frac{n!\,(a+b)_n}{(\mu+a)_n(\mu+b)_n}}\,\tilde{S}_n^\mu(y^2;a,b)$ and using $\frac{\Gamma(n+a)}{\Gamma(n+b)} \approx n^{a-b}$, $(z)_n = \frac{\Gamma(n+z)}{\Gamma(z)}$ and $a^{ib} = e^{ib\ln a}$ we obtain finally

$$S_n^\mu(y^2;a,b) \approx \frac{2\sqrt{\Gamma(\mu+a)\Gamma(\mu+b)\Gamma(a+b)}\,|\Gamma(2iy)|}{|\Gamma(a+iy)\Gamma(b+iy)\Gamma(\mu+iy)|\sqrt{n}}\cos(y\ln n - \gamma). \quad (A16)$$

Giving,

$$\sqrt{\rho^\mu(y;a,b)}\,S_n^\mu(y^2;a,b) \approx \sqrt{\frac{2}{n\pi}}\cos(y\ln n - \gamma), \quad (A17)$$

where $\gamma = \arg\{\Gamma(\mu+iy)\Gamma(a+iy)\Gamma(b+iy)/\Gamma(2iy)\}$. Similarly, and despite the fact that the continuous dual Hahn polynomials are a limiting case of the Wilson

–20–

polynomials, we cannot use Wilson's results in [26] simply because we cannot interchange this limit with the asymptotic limit.

Following a similar argument for obtaining bound states and resonances as in the Meixner-Pollaczek case and in the light of the asymptotic formula (A16), we conclude that bound states and resonances in the continuous dual Hahn case correspond to any one or more of the following three conditions

$$\left.\begin{array}{c}\mu+iy\\a+iy\\b+iy\end{array}\right\}=-n, \tag{A18}$$

In this case, however, (A18) leads to a finite number of bound states and only if the parameters $\mu$, $a$, or $b$ are negative. If we write (A18) as $c+iy=-n$, where $c$ stands for any one of the three parameters, then $n=0,1,..,n_{max}$, where $n_{max}$ is the largest integer less than or equal to $-c$. The following analysis gives justification to our conclusion:

The above analysis that resulted in (A16) is valid for $y>0$. Now, let us assume that $y$ is in the complex plane cut along $[0,\infty)$. First, we consider the case when the imaginary part of $iy$ is negative. In this case ${}_2F_1$ in (27) and (A10) converges at $t=1$ and we find that

$$S_n^\mu(y^2;a,b) \approx \frac{\sqrt{\Gamma(\mu+a)\Gamma(\mu+b)\Gamma(a+b)}\,\Gamma(-2iy)n^{-iy-1/2}}{\Gamma(a-iy)\Gamma(b-iy)\Gamma(\mu-iy)}. \tag{A19}$$

This shows that $\sum_{n=0}^{\infty}\left|S_n^\mu(y^2;a,b)\right|^2$ diverges for all $y$ with $\text{Re}(iy)$ negative when

$$\frac{1}{\Gamma(a-iy)\Gamma(b-iy)\Gamma(\mu-iy)} \neq 0. \tag{A20}$$

Thus, there are no bound states in this case. On the other hand, if $\text{Re}(iy)$ is positive, a similar analysis shows that (A19) remains valid with $iy \to -iy$ and we reach the same conclusion when

$$\frac{1}{\Gamma(a+iy)\Gamma(b+iy)\Gamma(\mu+iy)} \neq 0. \tag{A21}$$

Therefore, bound states occur at the zeros of $\Gamma(a\pm iy)\Gamma(b\pm iy)\Gamma(\mu\pm iy)$ when $\pm\text{Re}(iy)>0$, respectively. In that case, $\pm\text{Re}(iy)=-(n+c)>0$, where $c$ stands for any one of the three parameters. Thus, $c$ must be negative and $n$ can take only non-negative integer values that are less than or equal to $-c$. Moreover, one can also show that it does not matter which branch of the complex plane we use; if the parameters are positive we get no bound states.

**Table Caption:**

**Table 1**: A list of additional examples in the class of systems associated with potential functions showing the corresponding orthogonal polynomial, wavefunction, potential function and energy spectrum.

**Figures Caption:**

**Fig. 1**: Bound state and resonance energies (in units of $\alpha^2$) in the complex energy plane for the system whose energy spectrum formula is given by Eq. (54). We took $\beta = 0.9\alpha$. The figure shows the single bound state energy $E_0$ and several resonances that approach the origin as $n \to \infty$.

**Fig. 2**: Magnitude of the wavefunction $|\psi_m(x)|$ for the bound state ($m = 0$) and few of the lowest order resonances for the problem with the energy spectrum formula (54). We took $\beta = 0.9\alpha$ and $\phi_n^\lambda(x) = \sqrt{\frac{\Gamma(n+1)}{\Gamma(n+\nu+1)}} (\lambda x)^{\frac{\nu}{2}} e^{-\lambda x/2} L_n^\nu(\lambda x)$ where $\nu = 1$. The $x$-axis is in units of $\lambda^{-1}$.

**Fig. 3**: Resonance structure for the same problem as in Fig. 1 but with $\beta > \alpha$. The "bound-state-embedded resonances" are the two located in the third quadrant of the complex energy plane. We took $\alpha = 3.0$ and $\beta = 7.0$ (in atomic units).

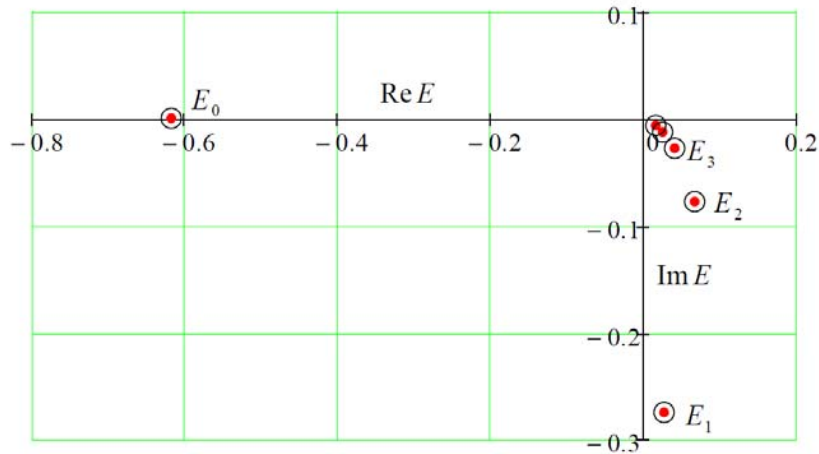

**Fig. 1**

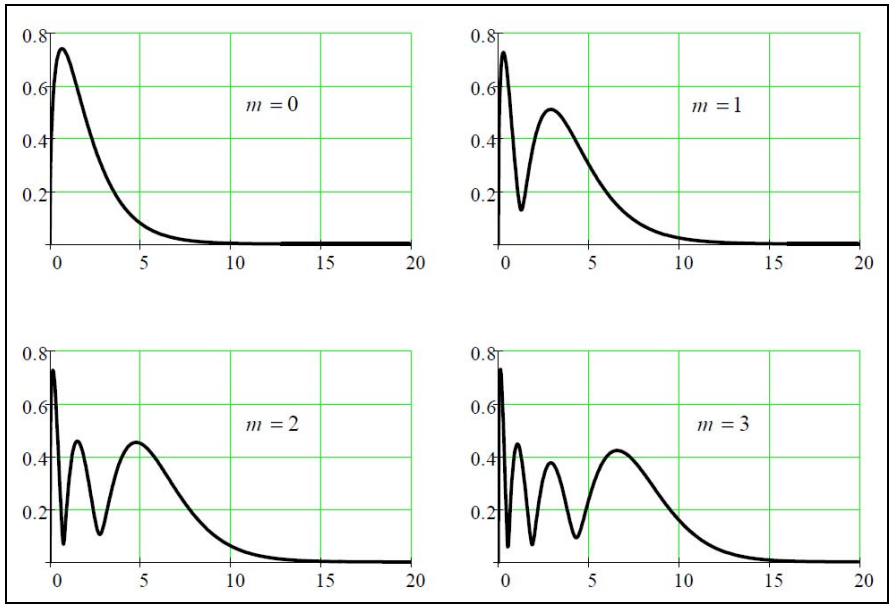

**Fig. 2**

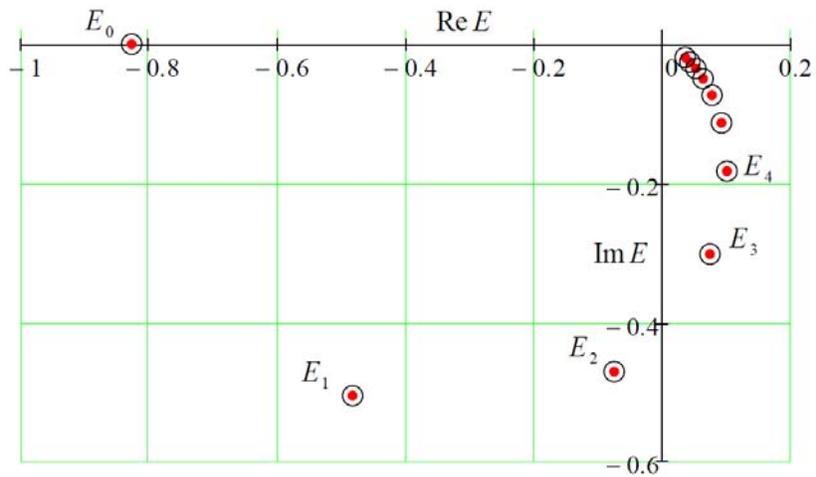

**Fig. 3**

## Table 1

| | Jacobi polynomial $P_n^{(\mu,\nu)}(y)$<br>$y \in [-1,+1]$, $\mu,\nu > -1$ | Bessel function $J_n(y)$<br>$n$ odd or even, $y \geq 0$ | Generalized Hermite polynomial $H_n^\mu(y)$<br>$\mu > -\frac{1}{2}$, $y \in [-\infty,+\infty]$ |
|---|---|---|---|
| Orthogonal Polynomial | | | |
| Weight Function $\omega(y)$ | $(1-y)^\mu (1+y)^\nu$ | $y^{-1}$ | $\|y\|^{2\mu} e^{-y^2}$ |
| Differential Equation | $\left\{(1-y^2)\dfrac{d^2}{dy^2} - [(\mu+\nu+2)y + \mu-\nu]\dfrac{d}{dy} + n(n+\mu+\nu+1)\right\} P_n^{(\mu,\nu)}(y) = 0$ | $\left[y^2 \dfrac{d^2}{dy^2} + y\dfrac{d}{dy} + y^2 - n^2\right] J_n(y) = 0$ | $\left[\dfrac{d^2}{dy^2} + 2\left(\dfrac{\mu}{y} - y\right)\dfrac{d}{dy} - \dfrac{\theta_n}{y^2} + 2n\right] H_n^\mu(y) = 0$<br>$\theta_{2n} = 0$, $\theta_{2n+1} = 2\mu$ |
| Potential Function $V(x)$ | $\dfrac{1}{4}\left(\dfrac{\pi}{a}\right)^2 \left[\dfrac{\mu^2+\nu^2-\frac{1}{2}}{\cos^2(\pi x/a)} + (\mu^2-\nu^2)\dfrac{\sin(\pi x/a)}{\cos^2(\pi x/a)}\right]$ | $\dfrac{n^2-1/4}{2x^2}$ | $\dfrac{\ell(\ell+1)}{2x^2} + \dfrac{1}{2}k^4 x^2$ |
| Wavefunction $\phi_n(x)$ | $\sqrt{(1-y)^{\mu+\frac{1}{2}}(1+y)^{\nu+\frac{1}{2}}}\, P_n^{(\mu,\nu)}(y)$,<br>$y = \sin(\pi x/a)$, $x \in \left[-\frac{a}{2}, +\frac{a}{2}\right]$ | $\sqrt{kx}\, J_n(kx)$, $n \geq 1$ | $\|kx\|^\mu e^{-k^2 x^2/2} H_n^\mu(kx)$,<br>$\mu = \begin{cases} \ell+1 & ,n \text{ even} \\ \ell & ,n \text{ odd} \end{cases}$ |
| Energy Spectrum $E_n$ | $\frac{1}{2}\left(\frac{\pi}{a}\right)^2 (n+\mu+\nu+1)^2$ | $\frac{1}{2}k^2$ | $k^2(2n+\ell+3/2)$ |